\documentclass[a4paper]{article}

\usepackage[pages=all, color=black, position={current page.south}, placement=bottom, scale=1, opacity=1, vshift=5mm]{background}
\SetBgContents{
	\tt    
}      

\usepackage[margin=1in]{geometry} 

\usepackage{amsmath}
\usepackage{amsthm}
\usepackage{amssymb}
\usepackage{subfig}
\usepackage{graphicx}

\usepackage[utf8]{inputenc}
\usepackage{hyperref}
\hypersetup{
	unicode,
	pdfauthor={Author One, Author Two, Author Three},
	pdftitle={A simple article template},
	pdfsubject={A simple article template},
	pdfkeywords={article, template, simple},
	pdfproducer={LaTeX},
	pdfcreator={pdflatex}
}

\newcommand{\be}{\begin{equation}}
\newcommand{\ee}{\end{equation}}
\newcommand{\bea}{\begin{eqnarray}}
\newcommand{\eea}{\end{eqnarray}}

\usepackage[sort&compress,numbers,square]{natbib}

\theoremstyle{plain}

\theoremstyle{definition}

\usepackage{graphicx, color}
\graphicspath{{fig/}}

\usepackage{algorithm, algpseudocode} 
\usepackage{mathrsfs} 

\usepackage{lipsum} 

\begin{document}
\title{Conformal Cyclic Cosmology from Varying Fundamental Constants}

\author{Konrad Marosek${^1} {^{\divideontimes}}$ \and Adam Balcerzak$^{2,3}$}

\date{
\small{$^1$Institute of Mathematics, Physics and Chemistry, Maritime University of Szczecin, Wa{\l }y Chrobrego 1-2, 70-500 Szczecin, Poland \\
$^2$Institute of Physics, University of Szczecin,  Wielkopolska 15, 70-451 Szczecin,  Poland \\
$^3$Copernicus Center for Interdisciplinary Studies, Szczepa\'nska 1/5, 31-011 Krak\'ow, Poland \\
$^{\divideontimes}$Corespondence: k.marosek@pm.szczecin.pl}}

\maketitle

\begin{abstract}
We investigate a bimetric cosmological model in which the gravitational and matter sectors are described by distinct metrics related through a time-dependent function $\alpha(t)$. This relation leads to dynamical gravitational parameters, with $G(t)\propto\alpha^{-3}$ and $c_{\rm grav}(t)\propto\alpha^{-1}$, where $c_{\rm grav}$ denotes the propagation speed of gravitational interactions. We consider a class of solutions for which $\alpha\rightarrow\infty$ at a finite cosmic time, resulting in $G\rightarrow0$ and $c_{\rm grav}\rightarrow0$. For an analytically tractable solution, we show that the final singularity occurs at finite conformal time and possesses a conformal structure compatible with a possible transition between successive cosmological cycles. The Tipler and Królak criteria indicate that the singularity is strong. We further discuss the physical consequences of the weakening gravitational interaction, including the dissolution of gravitationally bound structures and the shrinking of black-hole horizons. These effects suggest a possible mechanism leading toward an effectively radiation-dominated final state, which may be relevant for Conformal Cyclic Cosmology. The results provide a motivation for extending the analysis to more general scalar--tensor theories of gravity.
\end{abstract}

\section{Introduction}
\label{sec:intro}

In the framework of Conformal Cyclic Cosmology (CCC) \cite{Penrose1,Penrose2}, the Universe is envisioned as an endless sequence of cosmic epochs, known as aeons. Each aeon begins with a Big Bang and undergoes a prolonged period of expansion, eventually approaching a state that resembles a de Sitter universe. Rather than ending, this stage evolves toward a future conformal boundary, which can be represented as a three-dimensional spacelike hypersurface \cite{Penrose3}. Through a conformal transformation, this boundary becomes identified with the Big Bang of the next aeon, establishing a cyclic cosmological picture.

During the evolution of an aeon, matter gradually forms increasingly complex structures and is ultimately accumulated by black holes. As cosmic expansion continues, the Universe becomes progressively colder and more dilute. Once the ambient temperature falls below the effective Hawking temperature of black holes, they begin to lose mass through Hawking radiation \cite{Hawking1}. Over extremely long timescales, this process leads to the evaporation of all black holes, leaving a Universe dominated by radiation and effectively massless degrees of freedom. In this conformally invariant state, the distinction between very large and very small scales loses its physical significance, allowing the remote future of one aeon to be smoothly mapped onto the Big Bang of the next, thereby completing the cosmic cycle \cite{Penrose4}. 

In general relativity, spacetime singularities are commonly characterized by geodesic incompleteness, meaning that at least one timelike or null geodesic cannot be extended to arbitrary values of its affine parameter. This definition, established through the singularity theorems of Hawking and Penrose, provides a general mathematical criterion for identifying singular behavior in spacetime \cite{HawkingElis}. Nevertheless, geodesic incompleteness alone does not fully describe the physical properties of a singularity. As a result, additional classification schemes have been developed to distinguish between different singular states that may arise during the cosmological evolution of the Universe.

The discovery of a variety of novel cosmological singularities revealed important limitations of the traditional Hawking–Penrose characterization based solely on geodesic incompleteness. Many of these exotic singular scenarios exhibit divergences in physical or geometrical quantities while allowing timelike and null geodesics to be extended through the singular event. Consequently, geodesic behavior alone is often insufficient to distinguish between different classes of singularities. To address this issue, additional classification criteria have been introduced. In particular, the concepts developed by Tipler and Królak provide a measure of the strength of a singularity by examining the cumulative effect of tidal forces on physical objects approaching the singular region, thereby offering a more refined characterization of singular behavior in cosmological models \cite{Tipler,Królak}.

Another important tool in the study of cosmological singularities is provided by the energy conditions \cite{Curiel}. These conditions constitute a set of constraints imposed on the stress-energy tensor and are intended to encode the physically reasonable assumption that energy densities measured by observers should remain non-negative. The most frequently studied conditions are the null, weak, strong, and dominant energy conditions. Their fulfillment or violation plays a fundamental role in determining the causal and geometrical properties of spacetime and strongly influences the occurrence and nature of singular phenomena.

The discovery of the late-time accelerated expansion of the Universe \cite{Tonry,Knop,Tegmark} has significantly changed the discussion of energy conditions. Observational evidence suggests the presence of an exotic component with sufficiently negative pressure, commonly referred to as dark energy \cite{Steinghardt}. Among the proposed dark-energy candidates, phantom energy occupies a special position. Phantom matter is characterized by an equation-of-state parameter satisfying $ w<-1 $, which leads to the violation of several classical energy conditions, including the null and weak energy conditions \cite{Dabrowski1}. Such violations open the possibility of cosmological scenarios that are inaccessible within standard matter models and naturally lead to the emergence of novel singular behaviors.

These developments have motivated extensive studies of exotic cosmological singularities \cite{Haro}. Besides the standard Big-Bang singularity, a variety of future singular states have been identified, including the Big-Rip (Type I) \cite{Dabrowski1,Caldwell}, Sudden Future Singularity (Type II) \cite{Barrow1}, Finite Scale Factor Singularity (Type III) \cite{Odintsov1,Dabrowski2}, Big-Separation Singularity (Type IV) \cite{Odintsov2}, and the (w)-singularity (Type V) \cite{Dabrowski3}. Furthermore, scenarios such as the Little Rip and Pseudo Rip describe singularity-like evolution occurring asymptotically in the infinite future \cite{Frampton,Frampton1}. The classification of these singularities is typically based on their geodesic properties, singularity strength, and the fulfillment or violation of the classical energy conditions, providing a comprehensive framework for the investigation of possible cosmological futures.

The possibility that fundamental constants may vary was first considered by Weyl \cite{Weyl} and Eddington \cite{Eddington1,Eddington2}. Later, Dirac's Large Numbers Hypothesis \cite{Dirac} established a connection between cosmological and microscopic quantities, leading to the proposal that the gravitational constant may vary inversely with cosmic time. This idea motivated the development of the Brans-Dicke scalar-tensor theory of gravity \cite{BransDicke}, in which the effective gravitational coupling is determined by a scalar field and is proportional to its inverse.

Subsequently, a variety of models involving varying constants were proposed. These include theories with a varying speed of light $c$ \cite{Barrow2,AlbrechtMagueijo,BarrowMagueijo}, elementary charge $e$, proton-to-electron mass ratio $m_p/m_e$, and fine-structure constant $\alpha$ \cite{Bekenstein1,Bekenstein2}. Since many fundamental constants are related through physical laws, variations in one quantity may induce variations in others.

Theories with varying constants have been extensively investigated as possible alternatives to standard cosmological scenarios. In particular, they have been shown to address several outstanding problems of cosmology, including the flatness problem, the horizon problem, and the cosmological constant ($\Lambda$) problem \cite{Barrow2,AlbrechtMagueijo}.

The paper is organized as follows. Section \ref{sec2} introduces the bimetric framework considered in this work. In Section \ref{sec3}, we investigate the conformal structure of the model in the presence of a finite-time singularity. Section \ref{sec4} is devoted to the strength and physical interpretation of the singularity and its possible implications for the late-time cosmological evolution. Finally, Section \ref{sec5} summarizes the main results and concludes the paper.

\section{Bimetric model}
\label{sec2}

Cosmological models with distinct propagation speeds for matter and gravitational interactions can be naturally formulated within the framework of bimetric gravity \cite{Clayton}. In such theories, electromagnetic radiation propagates according to the causal structure defined by the matter metric, whereas the gravitational field is governed by a distinct gravitational metric. Consequently, the propagation speed of gravitational interactions, when measured with respect to the spacetime defined by the matter metric, may differ from the propagation speed associated with the matter sector.

Following this approach, we introduce two metrics: the gravitational metric ${\hat{g}}_{\mu \nu}$, which determines the causal structure of the gravitational sector, and the matter metric $g_{\mu \nu}$, which governs the dynamics of matter fields \cite{Marosek1}. Their relation is assumed to be
\be
\label{MetricForm1}
{\hat{g}}_{\mu \nu}
=
g_{\mu \nu}
{\left[
\alpha
-
\left(
\alpha-1
\right)
\left(
{\delta}_{0\mu}{\delta}_{0\nu}
\right)
\right]}^2~,
\ee
where $\alpha=\alpha(t)$ is a dimensionless function of cosmic time. This relation represents a particular disformal transformation and provides a specific realization of a bimetric theory.

For the diagonal cosmological background considered here, Eq. (\ref{MetricForm1}) implies that the temporal components of the two metrics coincide, whereas their spatial components are related by a factor $\alpha^2$:
\be
\label{g00}
{\hat{g}}_{00}=g_{00}~,
\ee
\be
\label{g11}
{\hat{g}}_{11}=\alpha^2 g_{11}~,
\ee
\be
\label{g22}
{\hat{g}}_{22}=\alpha^2 g_{22}~,
\ee
\be
\label{g33}
{\hat{g}}_{33}=\alpha^2 g_{33}~.
\ee

Thus, the spatial part of the gravitational metric is rescaled relative to the matter metric by the factor $\alpha^2$. Consequently, the cosmological expansion described in the matter frame generally differs from that defined by the gravitational frame.

Since electromagnetic radiation follows the causal structure of the matter metric, its propagation is determined by the matter frame. Gravitational waves, on the other hand, propagate according to the gravitational metric. The relation between the two metrics therefore allows the cosmological expansion to affect the propagation of electromagnetic and gravitational signals differently. This distinction is particularly relevant in the context of multimessenger astronomy, where observations of binary neutron-star mergers provide stringent constraints on the relative propagation of gravitational and electromagnetic signals \cite{Ligo1,Ligo2}.

The metric relation (\ref{MetricForm1}) can be embedded naturally within the framework of disformal gravity, in which two metrics are related through a diffeomorphism-invariant transformation involving a scalar field. The general form of such a transformation can be written as \cite{Bekenstein3}
\be
\label{DisformalMetric}
{\tilde{g}}_{\mu \nu}
=
C(\phi,X)g_{\mu \nu}
+
D(\phi,X)
{\partial}_{\mu}\phi{\partial}_{\nu}\phi~,
\ee
where
$X=\partial_\mu\phi\partial^\mu\phi$
denotes the kinetic term of the scalar field $\phi$, while
$C(\phi,X)$ and $D(\phi,X)$ are general functions of $\phi$ and $X$.
To recover the metric relation (\ref{MetricForm1}), we consider the particular choice
$C(\phi,X)\equiv\phi^2$
and
$D(\phi,X)=f(\phi)+g(\phi)\partial_\mu\phi\partial^\mu\phi$,
where $f(\phi)$ and $g(\phi)$ are unspecified functions of the scalar field.

We further restrict the analysis to a homogeneous and isotropic cosmological background described by the Friedmann--Lemaître--Robertson--Walker line element. Spatial homogeneity then requires the scalar field to depend only on cosmic time, i.e., $\phi=\phi(t)$. Combining this condition with the above choice of the functions $C(\phi,X)$ and $D(\phi,X)$ yields the following relations between the diagonal components of the two metrics:
\bea
\label{relg00}
{\hat{g}}_{00}
&=&
\phi^2 g_{00}
+
f(\phi)\dot{\phi}^2
+
g(\phi)\dot{\phi}^4 g^{00}~,
\\
\label{relgkk}
{\hat{g}}_{kk}
&=&
\phi^2 g_{kk}~.
\eea

Equation (\ref{relgkk}) reproduces the spatial relation between the two metrics given by Eqs. (\ref{g11})--(\ref{g33}). We now choose the time coordinate $x^0$ such that
$g_{00}=g^{00}=-1$, corresponding to the proper time of comoving observers. Requiring the temporal components of the two metrics to coincide,
${\hat{g}}_{00}=g_{00}$,
then imposes the condition
\be
\label{govphi}
-\phi^2
+
f(\phi)\dot{\phi}^2
-
g(\phi)\dot{\phi}^4
=
-1~.
\ee

The dynamics of the system is described by the total action \cite{Marosek1}
\be
\label{Action}
S=S_g[\hat{g}]+S_{\rm matter}[g]~.
\ee
The gravitational sector is governed by
\be
\label{gravaction}
S_g[\hat{g}]
=
-\frac{1}{16\pi G_0}
\int d^4x\,R[\hat{g}]\sqrt{-\hat{g}}~,
\ee
which corresponds to the Einstein--Hilbert action constructed from the gravitational metric $\hat{g}_{\mu\nu}$. The matter contribution is given by
\be
\label{fieldaction}
S_{\rm matter}[g]
=
-\frac{1}{2c_0}
\int d^4x\,L_{\rm matter}\sqrt{-g}~,
\ee
and depends exclusively on the matter metric $g_{\mu\nu}$. The constants $G_0$ and $c_0$ have the same dimensions as the gravitational constant $G$ and the speed of light $c$, respectively.

Variation of the total action (\ref{Action}) with respect to the metric $g^{\mu\nu}$ yields the corresponding field equations. Their temporal and spatial components can be written as \cite{Marosek1}
\bea
\label{voa00}
\alpha^3
\left(
R_{00}[\hat{g}]
-\frac{1}{2}\hat{g}_{00}R[\hat{g}]
\right)
&=&
\frac{8\pi G_0}{c_0^4}T_{00}~,
\\
\label{voa11}
\alpha
\left(
R_{ii}[\hat{g}]
-\frac{1}{2}\hat{g}_{ii}R[\hat{g}]
\right)
&=&
\frac{8\pi G_0}{c_0^4}T_{ii}~.
\eea

It is important to emphasize an aspect of the variational procedure. In the present framework, the function $\alpha$ is explicitly time dependent. Such a dependence may arise from Eq. (\ref{govphi}) if the model is interpreted as an effective description emerging from disformal gravity, with the identification $\phi\equiv\alpha$. For a prescribed time dependence $\alpha(t)=\phi(t)$, Eq. (\ref{govphi}) can be satisfied, at least locally, by an appropriate choice of the functions $f(\phi)$ and $g(\phi)$ subject to the constraint imposed by Eq. (\ref{govphi}). Consequently, within the effective description adopted here, variation of the action (\ref{Action}) with respect to the metric components $g^{\mu\nu}$ is sufficient to obtain the equations governing the cosmological dynamics considered below.

For a Friedmann--Lemaître--Robertson--Walker geometry in the matter frame, the metric is given by
\be
\label{FriedmanMetric}
ds_M^2
=
-c_0^2dt^2
+
a^2(t)
\left[
\frac{dr^2}{1-kr^2}
+
r^2
\left(
d\theta^2+\sin^2\theta\,d\varphi^2
\right)
\right]~,
\ee
where $k=0,\pm1$ denotes the spatial curvature parameter. The corresponding gravitational metric then takes the form
\be
\label{NewMetric}
ds_G^2
=
-c_0^2dt^2
+
\alpha^2(t)a^2(t)
\left[
\frac{dr^2}{1-kr^2}
+
r^2
\left(
d\theta^2+\sin^2\theta\,d\varphi^2
\right)
\right]~.
\ee

Substitution of Eqs. (\ref{FriedmanMetric}) and (\ref{NewMetric}) into the field equations (\ref{voa00}) and (\ref{voa11}), together with the assumption of spatial flatness, $k=0$, yields the energy density $\rho(t)$ and pressure $p(t)$:
\bea
\label{Density}
\rho(t)
=
\frac{3\alpha^3(t)}{8\pi G_0}
\left(
\frac{\dot{a}^2(t)}{a^2(t)}
+
\frac{2\dot{a}(t)\dot{\alpha}(t)}
{a(t)\alpha(t)}
+
\frac{\dot{\alpha}^2(t)}
{\alpha^2(t)}
\right)~,
\eea
\bea
\label{Pressure}
p(t)
=
-\frac{c_0^2\alpha(t)}{8\pi G_0}
\left(
\frac{\dot{a}^2(t)}{a^2(t)}
+
\frac{6\dot{a}(t)\dot{\alpha}(t)}
{a(t)\alpha(t)}
+
\frac{\dot{\alpha}^2(t)}
{\alpha^2(t)}
+
\frac{2\ddot{a}(t)}{a(t)}
+
\frac{2\ddot{\alpha}(t)}{\alpha(t)}
\right)~.
\eea

The associated continuity equation reads
\be
\label{ConEqu}
\dot{\rho}(t)
+
3\frac{\dot{a}(t)}{a(t)}
\left[
\rho(t)
+
\frac{\alpha^2(t)}{c_0^2}p(t)
\right]
+
3\frac{\dot{\alpha}(t)}{\alpha(t)}
\left[
\frac{\alpha^2(t)}{c_0^2}p(t)
\right]
=0~.
\ee

To establish consistency with the underlying disformal gravity framework, we identify the function $\alpha$ with the scalar field, $\alpha\equiv\phi$. With this identification, Eqs. (\ref{Density}) and (\ref{Pressure}) acquire a structure analogous to the corresponding field equations of the Brans--Dicke theory. Moreover, $\alpha$ becomes directly related to the effective gravitational coupling and the propagation speed of gravitational interactions. Within the present framework, these quantities are dynamical and are given by
\bea
\label{GravCoupling}
G(t)
=
\frac{G_0}{\alpha^3(t)}~,
\eea
\bea
\label{GravSpeed}
c_{\rm grav}(t)
=
\frac{c_0}{\alpha(t)}~.
\eea

In the limiting case of constant $\alpha$, Eqs. (\ref{Density}), (\ref{Pressure}), and (\ref{ConEqu}) reduce to the standard Friedmann equations. The corresponding effective values of the gravitational constant and the propagation speed of gravitational interactions are then
\bea
G_{\rm FM}
=
\frac{G_0}{\alpha^3}~,
\eea
and
\bea
c_{\rm FM}
=
\frac{c_0}{\alpha}~.
\eea

\section{Finite-time singularity and conformal structure}
\label{sec3}

Finite-time singularities can arise in general relativity and in a variety of modified theories of gravity. Their occurrence may be associated with matter fields that violate the standard energy conditions, such as phantom or ghost fields. A convenient parametrization of the scale factor describing the approach toward a finite-time singularity was introduced by Barrow \cite{Barrow2}:
\bea
\label{BarrowScaleFactor}
a_B \left( t \right) =
a_s \left[
\delta
+\left(1+\delta\right)
\left(\frac{t}{t_s}\right)^m
-\delta
\left(1-\frac{t}{t_s}\right)^n
\right],
\eea
where $\delta$, $a_s$, $n$, and $m$ are constants characterizing the cosmological evolution, while $t_s$ denotes the finite cosmic time at which the singularity occurs.

For the bimetric model considered in the previous section, we adopt the following parametrizations of the scale factor and of the function $\alpha(t)$:
\bea
\label{ScaleFactor4}
a(t)=a_0 t^m,
\eea
\bea
\label{AlphaFunction}
\alpha(t)=\left(1-\frac{t}{t_s}\right)^n.
\eea
These choices allow for a finite-time singularity associated with the vanishing of the effective gravitational coupling and of the propagation speed of gravitational interactions \cite{Marosek2}. For a spatially flat Universe ($k=0$), the conformal time associated with the gravitational metric is defined by
\bea
\label{conftime}
\eta = \int \frac{c_0\,dt}{a(t)\alpha(t)}.
\eea

We now consider the regime in which
$\alpha\rightarrow\infty$. According to the relations derived in the previous section,
this limit corresponds to
\bea
G(t)\rightarrow0,
\qquad
c(t)\rightarrow0,
\eea
where $c(t)$ denotes the effective propagation speed of gravitational interactions. Such a behavior is obtained for $n<0$. For the scale factor, we choose
\bea
a(t)=a_0 t^{1/2},
\eea
and consider the analytically tractable case $n=-1$. The corresponding conformal time can then be evaluated explicitly, giving
\bea
\label{ConformalTime}
\eta = \frac{2\sqrt{t}\left(3t_s-t\right)}{3a_0t_s}.
\eea

The initial Big-Bang state is reached in the limit
$t\rightarrow0$, for which
$\eta\rightarrow0$. The finite-time singularity is approached as
$t\rightarrow t_s^{-}$, corresponding to the finite conformal time
\bea
\label{ConformalTimeSingularity}
\eta_s =
\frac{4\sqrt{t_s}}{3a_0}.
\eea
Thus, both the initial state and the final singularity occur at finite values of the conformal time coordinate.

To investigate the conformal structure of the model near the final singularity, we consider the limit
$t\rightarrow t_s^{-}$. In this regime,
\bea
\label{AsymptoticAa}
a(t)\alpha(t)
\sim
\frac{a_0 t_s^{3/2}}{t_s-t}.
\eea
Consequently, the conformal-time differential behaves as
\bea
\label{AsymptoticConformalTime}
d\eta
=
\frac{dt}{a(t)\alpha(t)}
\sim
\frac{t_s-t}{a_0t_s^{3/2}}\,dt.
\eea
Integrating the above expression in the vicinity of $t_s$ gives
\bea
\label{TimeRelation}
t_s-t
\propto
\left(\eta_s-\eta\right)^{1/2}.
\eea
It follows that the conformally rescaled spatial scale factor behaves as
\bea
\label{ConformalScaleFactor}
a(\eta)\alpha(\eta)
=
\frac{1}{\Omega(\eta)}
\propto
\left(\eta_s-\eta\right)^{-1/2},
\eea
where $\Omega(\eta)$ denotes the corresponding conformal factor. This behavior indicates that the singularity can be represented as a conformal boundary of the spacetime. In particular, the conformal description provides a geometrical framework in which the evolution may be continued across the boundary and interpreted as a transition between successive cosmological cycles.

However, the above geometrical interpretation does not by itself establish the physical regularity of the transition. In particular, the energy-momentum tensor associated with the gravitational sector does not necessarily become traceless in the limit $\alpha\rightarrow\infty$. This raises the question of whether the matter content can be consistently matched to the conformal structure required for a transition between successive aeons.

To address this issue, it is necessary to examine the physical consequences of the limit $\alpha\rightarrow\infty$ and, in particular, the behavior of gravitationally bound structures and the strength of the resulting spacetime singularity. These aspects are considered in the following section.

\section{Physical interpretation of the $\alpha\rightarrow\infty$ singularity}
\label{sec4}
The strength of the singularity can be investigated using the criteria introduced by Tipler \cite{Tipler} and Królak \cite{Królak}. According to the Tipler criterion, a singularity is classified as strong if the double integral of the Ricci tensor contracted with the four-velocity diverges along the corresponding geodesic,
\bea
\label{Tipler}
\int_0^{\tau} d\tau' \int_0^{\tau'} d\tau'' \left| R_{\mu \nu}u^{\mu} u^{\nu} \right| \rightarrow \infty ~,
\eea
where $R_{\mu\nu}$ is the Ricci tensor, $u^\mu$ denotes the four-velocity of the geodesic observer, and $\tau$ is the proper time. For the comoving geodesics considered here, we take
\bea
{\hat{u}}^{\mu}=u^{\mu}=\left[-1,0,0,0\right]~,
\eea
so that
\bea
\label{RicciTensor}
R_{\mu \nu} u^{\mu} u^{\nu}=R_{00}~.
\eea
For the gravitational metric considered in this work, the corresponding component of the Ricci tensor is
\bea
\label{RicciTensor}
R_{00}=-\frac{3\left[2\dot{a}(t)\dot{\alpha}(t)+\ddot{a}(t)\alpha(t)+a(t)\ddot{\alpha}(t)\right]}{a(t)\alpha(t)}~.
\eea

For the solution considered below, with
\bea
a \left( t \right) \propto t^{1/2}
\eea
and $n=-1$, the function $\alpha(t)$ diverges as the singularity is approached,
\bea
\alpha \left( t \right) \propto {\left( t_s - t \right)}^{-1}~, ~~~~~ t \rightarrow t_s^{-}~.
\eea
Consequently, the leading contribution to Eq. (\ref{RicciTensor}) near the singularity behaves as
\bea
R_{00}\sim -\frac{6}{(t_s-t)^2}~.
\eea
Thus, the curvature experienced by a comoving observer becomes singular as $t\rightarrow t_s^-$. Substitution of this asymptotic behavior into Eq. (\ref{Tipler}) shows that the corresponding double integral diverges as the singularity is approached. The singularity is therefore classified as strong according to the Tipler criterion. An analogous analysis based on the Królak criterion leads to the same classification, providing an independent indication of the strong character of the singularity \cite{Marosek1}.

The strong nature of the singularity is particularly relevant for the physical interpretation of the limit $\alpha\rightarrow\infty$. In the present model, the effective gravitational coupling and the propagation speed of gravitational interactions are given by
\bea
G(t)&=&\frac{G_0}{\alpha^3(t)},\
c_{\rm grav}(t)=\frac{c_0}{\alpha(t)}.
\eea
Hence,
\bea
G(t)\rightarrow0,
\qquad
c_{\rm grav}(t)\rightarrow0
\qquad
\text{for}\qquad
\alpha(t)\rightarrow\infty.
\eea
The approach to the singularity is therefore accompanied by a simultaneous weakening and slowing of gravitational interactions. This behavior suggests a progressive disruption of gravitationally bound structures. On sufficiently large scales, gravitationally bound systems such as galaxies would become increasingly unstable, while on smaller scales planetary systems could lose their gravitational binding. Similarly, the decreasing gravitational coupling would modify the balance between gravitational compression and internal pressure in stars, potentially leading to their expansion and subsequent dispersal. As a result, matter would become progressively diluted as the singularity is approached.

A similar argument can be applied to black holes. If the Schwarzschild radius is expressed in terms of the effective gravitational coupling and the gravitational interaction speed, one obtains
\bea
r_{\rm S}(t)=\frac{2G(t)M}{c_{\rm grav}^2(t)}
=\frac{2G_0M}{c_0^2\alpha(t)}~,
\eea
for a black hole of mass $M$. Therefore,
\bea
r_{\rm S}(t)\rightarrow0
\qquad
\text{as}
\qquad
\alpha(t)\rightarrow\infty.
\eea
Within this interpretation, black-hole horizons would progressively shrink as the singularity is approached. This behavior may provide a mechanism through which matter contained in compact gravitational structures is ultimately transferred to radiation. Thus, one may envisage a possible sequence in which gravitationally bound matter becomes progressively more compact, potentially forming black holes, whose horizons subsequently shrink as the effective gravitational coupling tends to zero. Their eventual evaporation could then provide a mechanism for converting the remaining matter into radiation, leaving a radiation-dominated state as the singularity is reached.

In this picture, the strong tidal effects associated with the Tipler and Królak singularity criteria describe the geometric destruction of extended structures, while the simultaneous limit $G\rightarrow0$ suppresses the gravitational binding responsible for maintaining such structures. The two effects therefore act in a mutually reinforcing manner: the strong curvature effects destroy extended geodesic volumes, whereas the vanishing gravitational coupling prevents the formation of a persistent gravitationally bound configuration.

Consequently, the limit
\bea
t \rightarrow t_s^{-}~,
\eea
may admit an interpretation in which the matter content becomes increasingly diluted and the remaining energy is progressively transferred into radiation. In the limiting state, the conventional notion of a gravitationally connected spacetime may cease to be applicable, as the effective gravitational interaction between spatially separated regions tends to zero. This provides a possible mechanism for obtaining a radiation-dominated final state, which is of particular interest in the context of conformal cyclic cosmology.

It should be emphasized, however, that the above interpretation involves an extrapolation beyond the regime in which classical gravitational theory can be expected to provide a complete description. In particular, the behavior of black-hole horizons and Hawking evaporation when both $G$ and $c_{\rm grav}$ become dynamical and approach zero cannot be established from the standard semiclassical treatment alone. A complete description of the final state would ultimately require a consistent theory of quantum gravity. Therefore, the proposed conversion of matter into radiation should be regarded as a possible physical interpretation of the limiting behavior of the model rather than as a direct consequence of the classical field equations.

\section{Conclusions}
\label{sec5}

In this work, we have investigated a bimetric cosmological model in which the gravitational and matter sectors are described by distinct metrics. The relation between the two metrics introduces a time-dependent function $\alpha(t)$, which in turn leads to dynamical effective values of the gravitational constant and the propagation speed of gravitational interactions. We have considered a particular class of solutions for which $\alpha$ diverges at a finite cosmic time, resulting in
\bea
G(t)\rightarrow 0,
\qquad
c(t)\rightarrow 0,
\eea
as $t\rightarrow t_s$.

For the specific solution considered in this work, the evolution of the Universe terminates at a finite value of the cosmic time. At the same time, the corresponding conformal time also approaches a finite limit $\eta_s$. The resulting conformal structure provides a possible geometrical setting for connecting the final state of one cosmological cycle with the initial state of a subsequent one. In this sense, the model offers a framework in which a conformal transition of the type required by Conformal Cyclic Cosmology (CCC) may occur at a finite cosmic time.

An important aspect of the model is the behavior of the Universe in the limit $\alpha\rightarrow\infty$. As the effective gravitational interaction weakens and its propagation speed approaches zero, gravitationally bound structures are expected to progressively lose their binding. This provides a possible mechanism through which matter could be driven toward a state in which gravitationally bound structures cease to exist. Within the classical interpretation adopted here, the decreasing gravitational constant also leads to a shrinking Schwarzschild radius, suggesting that black holes may lose their horizons as the singularity is approached.

These considerations indicate a possible route toward a final state dominated by radiation, which is particularly relevant in the context of CCC. A major conceptual difficulty in the original formulation of CCC is the requirement that the late-time Universe becomes effectively free of massive particles, since their presence prevents the conformal rescaling required to connect successive aeons. Penrose has therefore considered the possibility that massive matter ultimately disappears through processes such as the decay of massive particles. The mechanism suggested by the present model is qualitatively different: the weakening of the gravitational interaction may lead to the dissolution of gravitationally bound structures and, together with the shrinking of black-hole horizons, potentially drive the Universe toward a state in which radiation becomes the dominant surviving component.

The strength of the singularity was also examined using the Tipler and Królak criteria. For the solutions considered here, both criteria indicate a strong singularity. This result emphasizes that the final state cannot be regarded simply as a regular endpoint of the classical cosmological evolution. Rather, the singular behavior should be interpreted as a limiting regime in which the classical description may eventually cease to be sufficient.

Although the present model provides a possible connection between finite-time gravitational singularities and the conformal structure required by CCC, several aspects remain open. In particular, the ultimate fate of matter and black holes in the limit $\alpha\rightarrow\infty$ cannot be established rigorously within classical gravity alone. A complete description of this regime would likely require a quantum theory of gravity. Moreover, the present analysis is based on a specific bimetric construction and a particular parametrization of $\alpha(t)$ and $a(t)$. It is therefore important to investigate whether similar behavior occurs in more general scalar--tensor theories and other modified theories of gravity.

The results presented here suggest that varying gravitational parameters may provide an interesting perspective on the transition between successive cosmological aeons. Rather than assuming that the late-time Universe becomes radiation dominated solely through the decay of massive particles, the present framework suggests that the weakening of the gravitational interaction itself may contribute to the removal of massive, gravitationally bound structures. Further investigation of this possibility, particularly within scalar--tensor models, is left for future work.

\textbf{Acknowledgments} The authors would like to acknowledge the use of OpenAI's ChatGPT for assistance with language editing, paraphrasing, and improving the clarity and organization of the manuscript. All scientific content, calculations, interpretations, and conclusions were developed and verified by the authors.

\end{document}